\documentclass[12pt,twoside]{article}

\evensidemargin=\oddsidemargin
\def\Title#1#2#3{%
    \baselineskip=18pt
    \begin{center}
          {\Large\bf\uppercase{#1} \\ }
          \bigskip\bigskip
          {#2} \\
          {#3} \\
    \end{center}}

\long\def\Abstract#1{%
         \bigskip
         \parbox{0.93\textwidth}{%
                 \begin{center}
                       {\bf Abstract} \\
                 \end{center}
                 \medskip{\baselineskip=14pt #1}
                 \vss}
         \bigskip}

\begin{document}

\Title{QUANTUM GEOMETRODYNAMICAL\\
DESCRIPTION OF THE UNIVERSE\\
IN DIFFERENT REFERENCE FRAMES}%
{T. P. Shestakova}%
{Department of Theoretical and Computational Physics, Rostov State University, \\
Sorge Str. 5, Rostov-on-Don 344090, Russia \\
E-mail: {\tt stp@phys.rnd.runnet.ru}}

\Abstract{Several years ago the so-called quantum geometrodynamics in
extended phase space was proposed. The main role in this version of quantum
geometrodynamics is given to a wave function that carries information
about geometry of the Universe as well as about a reference frame
in which this geometry is studied. We consider the evolution of a
physical object (the Universe) in ``physical'' subspace of extended
configurational space, the latter including gauge and ghost degrees of
freedom. A measure of the ``physical'' subspace depends on a chosen
reference frame, in particular, a small variation of a gauge-fixing
function results in changing the measure. Thus, a transition to
another gauge condition (another reference frame) leads to
non-unitary transformation of a physical part of the wave function.
From the viewpoint of the evolution of the Universe in the ``physical''
subspace a transition to another reference frame is an irreversible
process that may be important when spacetime manifold has a nontrivial
topology.}

\section{Introduction}
Recently a new version of quantum geometrodynamics was proposed
\cite{Our1, Our2}. While constructing this new version a special
attention was paid to the fact that the Universe as a whole may not
possess asymptotic states in which one can separate the so-called
``non-physical'' degrees of freedom from physical ones. In this
report I shall refer to the case of a closed universe, but the same
situation is expected to be in a general case if the Universe has
some nontrivial topology.

In the path integral approach, which had been chosen to be a basic
tool in our investigation, the lack of asymptotic states makes us
refuse imposing asymptotic boundary conditions in a path integral.
It leads to a gauge-dependent wave function of the Universe that
satisfies a gauge-dependent Schr\"odinger equation. Indeed, in the
modern quantum theory of gauge fields it is these very asymptotic
boundary conditions that enable us to prove independence of a path
integral on a chosen gauge (see, for example, \cite{Hennaux}).

The goal of this report is to discuss some consequences of the
proposed formulation of quantum geometrodynamics concerning the
description of the Universe in different reference frames.

\section{The new formulation of quantum geometrodynamics: basic
equations}
To make the things more clear, let us turn to a simple
minisuperspace model with the gauged action
\begin{equation}
\label{action}
S=\!\int\!dt\,\biggl\{
  \displaystyle\frac12 v(\mu, Q^a)\gamma_{ab}\dot{Q}^a\dot{Q}^b
  -\frac1{v(\mu, Q^a)}U(Q^a)
  +\pi\left(\dot\mu-f_{,a}\dot{Q}^a\right)
  -i w(\mu, Q^a)\dot{\bar\theta}\dot\theta\biggr\}.
\end{equation}
Here $Q^a$ stands for physical variables such as a scale factor or
gravitational-wave degrees of freedom and material fields, and
we use an arbitrary parametrization of a gauge variable $\mu$
determined by the function $v(\mu, Q^a)$. For example, in the case
of isotropic universe or the Bianchi IX model $\mu$ is bound to the
scale factor $r$ and the lapse function $N$ by the relation
\begin{equation}
\label{paramet}
\displaystyle\frac{r^3}{N}=v(\mu, Q^a).
\end{equation}
$\theta,\,\bar\theta$ are
the Faddeev -- Popov ghosts after replacement
$\bar\theta\to -i\bar\theta$. Further,
\begin{equation}
\label{w_def}
w(\mu, Q^a)=\frac{v(\mu, Q^a)}{v_{,\mu}};\quad
v_{,\mu}\stackrel{def}{=}\frac{\partial v}{\partial\mu}.
\end{equation}
We confine attention to the special class of gauges not depending
on time
\begin{equation}
\label{frame_A}
\mu=f(Q^a)+k;\quad
k={\rm const},
\end{equation}
which can be presented in a differential form,
\begin{equation}
\label{diff_form}
\dot{\mu}=f_{,a}\dot{Q}^a,\quad
f_{,a}\stackrel{def}{=}\frac{\partial f}{\partial Q^a}.
\end{equation}

The Schr\"odinger equation for this model reads
\begin{equation}
\label{SE1}
i\,\frac{\partial\Psi(\mu,Q^a,\theta,\bar\theta;\,t)}{\partial t}
 =H\Psi(\mu,\,Q^a,\,\theta,\,\bar\theta;\,t),
\end{equation}
where
\begin{equation}
\label{H}
H=-\frac i w\frac{\partial}{\partial\theta}
   \frac{\partial}{\partial\bar\theta}
  -\frac1{2M}\frac{\partial}{\partial Q^{\alpha}}MG^{\alpha\beta}
   \frac{\partial}{\partial Q^{\beta}}
  +\frac1v(U-V);
\end{equation}
$M$ is the measure in the path integral,
\begin{equation}
\label{M}
M(\mu, Q^a)=v^{\frac K2}(\mu, Q^a)w^{-1}(\mu, Q^a);
\end{equation}
\begin{equation}
\label{Galpha_beta}
G^{\alpha\beta}=\frac1{v(\mu, Q^a)}\left(
 \begin{array}{cc}
  f_{,a}f^{,a}&f^{,a}\\
  f^{,a}&\gamma^{ab}
 \end{array}
 \right);\quad
\alpha,\beta=(0,a);\quad
Q^0=\mu,
\end{equation}
$K$ is a number of physical degrees of freedom; the wave function is defined
on extended configurational space with the coordinates
$\mu,\,Q^a,\,\theta,\,\bar\theta$.
$V$ is a quantum correction to the potential $U$, that depends on the chosen
parametrization (\ref{paramet}) and gauge (\ref{frame_A}):
\begin{eqnarray}
V&=&\frac5{12w^2}\left(w^2_{,\mu}f_{,a}f^{,a}+2w_{,\mu}f_{,a}w^{,a}
    +w_{,a}w^{,a}\right)
   +\frac1{3w}\left(w_{,\mu,\mu}f_{,a}f^{,a}+2w_{,\mu,a}f^{,a}
    +w_{,\mu}f_{,a}^{,a}+w_{,a}^{,a}\right)+\nonumber\\
&+&\frac{K-2}{6vw}\left(v_{,\mu}w_{,\mu}f_{,a}f^{,a}+v_{,\mu}f_{,a}w^{,a}
    +w_{,\mu}f_{,a}v^{,a}+v_{,a}w^{,a}\right)-\nonumber\\
&-&\frac{K^2-7K+6}{24v^2}\left(v^2_{,\mu}f_{,a}f^{,a}+2v_{,\mu}f_{,a}v^{,a}
    +v_{,a}v^{,a}\right)+\nonumber\\
\label{V}
&+&\frac{1-K}{6v}\left(v_{,\mu,\mu}f_{,a}f^{,a}+2v_{,\mu,a}f^{,a}
    +v_{,\mu}f_{,a}^{,a}+v_{,a}^{,a}\right).
\end{eqnarray}

Ones we agreed that imposing asymptotic boundary conditions is not
correct in the case of a closed Universe, we {\it are doomed} to come
to a gauge-dependent description of the Universe. The Schr\"odinger
equation (\ref{SE1}) -- (\ref{V}) is {\it a direct mathematical
consequence} of a path integral with the effective action (\ref{action})
without asymptotic boundary conditions, it is derived from the latter by
the standard well-definite Feynman procedure. Any additional
conditions like the requirement of BRST-invariance cannot help to
reduce Eq. (\ref{SE1}) to a gauge-invariant equation. So the
role of such conditions is questionable when one deals with a
system without asymptotic states.

\section{The description of quantum Universe in different reference frames}
The general solution to the Schr\"odinger equation (\ref{SE1}) has the
following structure:
\begin{equation}
\label{GS-A}
\Psi(\mu,\,Q^a,\,\theta,\,\bar\theta;\,t)
 =\int\Psi_k(Q^a,\,t)\,\delta(\mu-f(Q^a)-k)\,(\bar\theta+i\theta)\,dk.
\end{equation}
The dependence of the wave function (\ref{GS-A}) on ghosts is
determined by the demand of norm positivity.

Note that the general solution (\ref{GS-A}) is a
superposition of eigenstates of a gauge operator,
\begin{equation}
\label{k-vector}
\{\mu-f(Q^a)\}|k\rangle=k\,|k\rangle;\quad
|k\rangle=\delta\left(\mu-f(Q^a)-k\right).
\end{equation}
It can be interpreted in the spirit of Everett's ``relative state''
formulation. In fact, each element of the superposition (\ref{GS-A})
describe a state in which the only gauge degree of freedom $\mu$ is
definite, so that time scale is determined by processes
in the physical subsystem through functions
$v(\mu,\,Q^a),\,f(Q^a)$ (see (\ref{paramet}), (\ref{frame_A}) ),
while $k$ being determined initial clock setting.
The function $\Psi_k(Q^a,\,t)$ describes a state of the physical
subsystem for a reference frame fixed by the condition
(\ref{frame_A}). It is a solution to the equation
\begin{equation}
\label{phys.SE}
i\,\frac{\partial\Psi_k(Q^a;\,t)}{\partial t}
 =H_{(phys)}[f]\Psi_k(Q^a;\,t),
\end{equation}
\begin{equation}
\label{phys.H-A}
H_{(phys)}[f]=\left.\left[-\frac1{2M}\frac{\partial}{\partial Q^a}
  \frac1v M\gamma^{ab}\frac{\partial}{\partial Q^b}
 +\frac1v (U-V)\right]\right|_{\mu=f(Q^a)+k}.
\end{equation}

The peculiarity of this consideration is that a measure in the
subspace of physical degrees of freedom depends on a chosen gauge
condition. Indeed, the measure (\ref{M}) in the path integral is
proportional to a square root of the determinant of metric of
``physical'' configurational subspace, the latter depending on the
gauge variable $\mu$: $G^{phys}_{ab}=v(\mu, Q^a)$. So we get
$$
\int\Psi^*_{k'}(Q^a,\,t)\,\Psi_k(Q^a,\,t)\,
 \delta(\mu-f(Q^a)-k')\,\delta(\mu-f(Q^a)-k)\,
 dk'\,dk\,M(\mu,\,Q^a)\,d\mu\,\prod_adQ^a=
$$
\begin{equation}
\label{Psi_norm}
=\int\Psi^*_k(Q^a,\,t)\,\Psi_k(Q^a,\,t)\,
 M(f(Q^a)+k,\,Q^a)\,\prod_adQ^a\,dk=1.
\end{equation}

It is easy to see that a transition to another gauge condition
(another reference frame) cannot be described by an unitary
transformation of the physical part of the wave function
$\Psi_k(Q^a,\,t)$. As a consequence of this structure of physical
subspace, we will obtain different physical results in different
reference frames.

One can seek the solution to Eq.(\ref{phys.SE}) in the form
of superposition of stationary state eigenfunctions:
\begin{equation}
\label{stat.states}
\Psi_k(Q^a,\,t)=\sum_n c_{kn}\psi_n(Q^a)\exp(-iE_n t);\quad
H_{(phys)}[f]\psi_n(Q^a)=E_n\psi_n(Q^a).
\end{equation}

The parameter $E$ should not be associated with energy of any
material field. It is a new integral of motion that emerges in the
proposed formulation as a result of fixing a gauge condition and
characterizes a subsystem which corresponds to observation means
-- a reference frame (see \cite{Our1, Our2} for details).

The proposed formulation of quantum geometrodynamics suffers from
the fact that, according to it, the Universe could be created in any
state with a nonzero value of the parameter $E$. On the other hand,
we can surely say that at the present stage of its evolution the
Universe is found in the state with $E=0$. For this state only
we can obtain a gauge-invariant classical limit.
So we need some mechanism of the ``reduction'' of the wave function
to the state with $E=0$.

We also do not know a criterion for a choice of a reference frame.
While we do not know any deeper reason, our choice may be dictated by
convenience and simplicity. However, it is rather rare situation when
a spacetime manifold can be covered by only one coordinate system.
In a case of nontrivial topology spacetime may consist of
several regions covered by different coordinate charts. It is
a serious problem for the Dirac canonical quantization which
requires to introduce a foliation by spacelike hypersurfaces that
would cover all available spacetime. The approach presented here
which is based on path integration may turn to be more adequate
for description of this situation, since the path integral admits
(at least formally) introducing different gauge conditions in
different spacetime regions.

Then, in this approach, the wave function may satisfy different
Schr\"odinger equations in different spacetime regions, the form
of the equation in each region is fixed by a chosen reference frame.
Like a transition to another reference frame in the same spacetime
region, a transition from one spacetime region to another must be
described by a non-unitary transformation of the wave function.

The problem arises if it is possible to give a mathematical
description to the transition to a different reference frame. We
can try to do it for our minisuperspace model where the only role
of a gauge condition is in fixing time scale.

We now consider the equation for the physical part of the wave function
when varying the gauge-fixing function $f(Q^a)$. Let a reference
frame be fixed by the condition
\begin{equation}
\label{frame_B}
\mu=f(Q^a)+\delta f(Q^a)+k.
\end{equation}
We can choose a basis corresponding to this reference frame, so
that
\begin{equation}
\label{GS-B}
\Psi(\mu,\,Q^a,\,\theta,\,\bar\theta;\,t)
 =\int\tilde\Psi_k(Q^a,\,t)\,\delta(\mu-f(Q^a)-\delta f(Q^a)-k)\,
  (\bar\theta+i\theta)\,dk.
\end{equation}
The function $\tilde\Psi_k(Q^a,\,t)$ satisfies Eq.(\ref{phys.SE})
with a Hamiltonian
\begin{equation}
\label{phys.H-B}
H_{(phys)}[f+\delta f]=\left.\left[-\frac1{2M}\frac{\partial}{\partial Q^a}
  \left(\frac1v M\gamma^{ab}\frac{\partial}{\partial Q^b}\right)
 +\frac1v (U-V)\right]\right|_{\mu=f(Q^a)+\delta f(Q^a)+k}.
\end{equation}
If the variation of the gauge-fixing function $\delta f(Q^a)$ is small,
one can write
\begin{equation}
\label{phys.H-B1}
H_{(phys)}[f+\delta f]=H_{(phys)}[f]+W[\delta f]+V_1[\delta f]
\end{equation}
For our minisuperspace model the operator $W[\delta f]$ reads
\begin{eqnarray}
W[\delta f]&=&\left[\frac1{2M^2}\frac{\partial M}{\partial\mu}\delta f
 \frac{\partial}{\partial Q^a}\left(\frac1v M\gamma^{ab}
 \frac{\partial}{\partial Q^b}\right)\right.-\nonumber\\
\label{W}
&-& \left.\left.\frac1{2M}\frac{\partial}{\partial Q^a}
 \left(\left(\frac1v\frac{\partial M}{\partial\mu}
  -\frac M{v^2}\frac{\partial v}{\partial\mu}\right)
 \delta f\gamma^{ab}\frac{\partial}{\partial Q^b}\right)\right]
 \right|_{\mu=f(Q^a)+k},
\end{eqnarray}
and $V_1[\delta f]$ is the change of quantum potential $V$ in
first order of $\delta f$.

We can inquire how the probabilities of states (\ref{stat.states})
change under the perturbation $W[\delta f]+V_1[\delta f]$,
which is due to a small variation of the gauged-fixing function $f(Q^a)$.
The Hamiltonian (\ref{phys.H-B}) is Hermitian by construction in
a space with the measure $M(f(Q^a)+\delta f(Q^a)+k,\,Q^a)$,
however it is not Hermitian in a space with the measure
$M(f(Q^a)+k,\,Q^a)$ in which the functions (\ref{stat.states}) are
normalized. In this space the operator (\ref{W}) will
have, in general, anti-Hermitian part. So it follows already from
Eqs.\,(\ref{phys.H-B}) -- (\ref{W}) that a transition to
another reference frame is {\it an irreversible process}.

This conclusion is in accordance with our interpretation of the
reference frame as the only measuring instrument representing the
observer in quantum theory of gravity. The variation of a
gauge-fixing function means changing interaction with the measuring
instrument that implies an unremoval influence on properties of
the physical object.

\end{document}